# *FSCsec*: Collaboration in Financial Sector Cybersecurity - Exploring the Impact of Resource Sharing on IT Security

Sayed Abu Sayeed
*College of Business*
*Florida Atlantic University*
Boca Raton, FL, USA
ssayeed2024@fau.edu

Mir Mehedi Rahman
*School of Business & Technology*
*Emporia State University*
Emporia, KS, USA
mrahman2@g.emporia.edu

Samiul Alam
*School of Computing*
*University of South Alabama*
Mobile, AL, USA
sa2426@jagmail.southalabama.edu

Naresh Kshetri
*Department of Cybersecurity*
*Rochester Institute of Technology*
Rochester, NY, USA
naresh.kshetri@rit.edu

*Abstract*— The financial sector's dependence on digital infrastructure increases its vulnerability to cybersecurity threats, requiring strong IT security protocols with other entities. This collaboration, however, is often identified as the most vulnerable link in the chain of cybersecurity. Adopting both symbolic and substantive measures lessens the impact of IT security spending on decreasing the frequency of data security breaches in the long run. The Protection Motivation Theory clarifies actions triggered by data sharing with other organizations, and the Institutional theory aids in comprehending the intricate relationship between transparency and organizational conduct. We investigate how things like regulatory pressure, teamwork among institutions, and people's motivations to protect themselves influence cybersecurity. By using simple theories to understand these factors, this research aims to provide insights that can help financial institutions make better decisions to protect. We have also included the discussion, conclusion, and future directions in regard to collaboration in financial sector cybersecurity for exploring impact of resource sharing.

## I. Introduction

In today's interconnected digital world, cybersecurity remains a critical concern for financial institutions. The financial sector faces an ever-evolving threat landscape, with cyberattacks becoming more sophisticated and frequent. As a result, organizations must adopt robust measures to safeguard their systems, data, and customer information. The impact of information technology (IT) on a financial company's performance is a topic of interest for both practical applications and research. Sharing information between financial companies is becoming more common nowadays. Usually, sharing information is good for everyone involved, but sometimes keeping secrets is important. Banks are especially worried about cyber-attacks, so they're thinking about ways to work together better to stay safe online. Therefore, and not surprisingly, financial organizations often form strategic cooperative partnerships between banks and fintech [1]. Given the risk of financial institutions sharing important information with competitors, our research question is as follows:

"What factors contribute to reducing data breaches through collaborative resource sharing across financial institutions?"

The Sarbanes-Oxley Act makes sure companies have good controls inside and share information well [2]. This means accounting is even more important for handling information securely and figuring out if it's worth it. How companies share resources and invest in security depends on how they rely on each other and how much security costs them. While past studies have looked at how sharing information is important for businesses and IT security, no one has looked at how sharing resources affects financial institutions. Also, there are not many theories explaining how financial institutions sharing security information affects their success or motivates them to share. Before examining the details of our study, we first examine the magnitude of the IT security problem that drives this research.

Research shows that when companies have security breaches in their IT systems, it has a significant adverse influence on the overall performance of a company. Cyberattacks on banks are happening more often, and they're getting more advanced and causing bigger problems. These attacks are usually done by organized groups who want to get important information and disrupt the entire business. With the increasing digitization of banking and reliance on remote services, cyber threats have become more intense. The COVID-19 pandemic has made it even worse, with more attacks happening, some even sponsored by governments. Banks are especially at risk of ransomware attacks, where hackers lock up their systems until they pay up. These attackers don't just want money; they want to stay hidden and mess with lots of different systems. Banks have to find a balance between keeping things safe and making it easy for people to use their services online. To do that, it's important to understand why these attacks happen and how they work so we can stop them and keep our financial systems safe [3]. Another emerging threat to the banking industry is crypto-jacking. Crypto-jacking is the violation of using another computer and the features of it to dig out cryptocurrency. Crypto-jacking used to sneak around in web browsers, but now it's getting into important servers [4].

Financial institutions sometimes struggle to protect themselves from cyber-attacks because they don't have enough resources. This is especially true for smaller businesses that might not have the money or know-how to set up strong security measures. Even though basic security is important, not many companies are asking for it, which means there's not much push for improvement. While the government and insurance companies can help to a certain extent, there's still a shortage of people who possess the potential of how to keep things safe online. Additionally, working with other countries to fight cybercrime can be tricky because of different rules and regulations of doing things. Despite all this, it is important for financial institutions to focus on cybersecurity to stay safe from hackers and violations [5].

Financial organizations often team up and share resources to implement policies around cybersecurity and other regulatory issues. They do this through alliances, groups, and by working with government agencies. Organizations such as the Financial Services Information Sharing and Analysis Center (FS-ISAC) help financial institutions share information and work together to collectively address cybersecurity threats and shape policy creation. Furthermore, financial institutions can engage in regulatory forums and consultations, where they can provide their expertise and opinions to influence policy decisions. But we don't know much about how well these teamwork efforts work or what they mean. Integrating Protection Motivation Theory (PMT), Institutional Theory, and the concepts of Symbolic and Substantive Adoption could enrich the research. This could give us a deeper understanding of the topic.

Individuals' reactions to threats can be understood through the Protection Motivation Theory (PMT) [6]. When organizations team up for cybersecurity, they assess how serious and likely cyber-attacks are. Institutional theory looks at how organizations follow outside rules and norms [7]. Collaborations often happen because of these outside pressures, like rules or standards. Symbolic adoption means just doing things on the surface to look good or follow the law; on the other hand, substantive adoption means really making changes in how the organization works [8].

By using these theoretical perspectives in the research framework, researchers can provide a deeper analysis of the collaboration in cybersecurity within the financial sector. This analysis will not only focus on the technical aspects but also consider the institutional and motivational factors that influence collaborative efforts and their effects on IT security.

## II. COLLABORATIVE EFFORTS AND THEORETICAL FRAMEWORK FOR COLLABORATION

In recent years, the financial sector has faced an increasing number of cybersecurity threats, particularly due to its reliance on digital infrastructure. This dependence makes financial institutions vulnerable to cyberattacks, emphasizing the need for effective collaboration and resource sharing to bolster IT security. Various initiatives and partnerships have been established to mitigate these risks. Public-private partnerships in the United States, such as the Financial Services Information Sharing and Analysis Center (FS-ISAC), are examples of successful collaborations that foster information sharing and joint response strategies, ultimately enhancing the sector's resilience against cyber threats. Global efforts also play a significant role, especially in addressing cybersecurity challenges in developing nations. Industry-led initiatives further underscore the importance of collaboration between financial institutions, regulators, and third-party vendors. The table below (Table 1) provides an overview of the key areas of collaboration in the financial sector's cybersecurity efforts.

The foundation of collaboration in financial sector cybersecurity is supported by several key theories that explain why and how institutions cooperate to enhance their security posture.

TABLE I.

| Category | Details |
|---|---|
| Public-Private Partnerships | In the United States, the Financial Services Sector (FSS) is often seen as a leading example of public-private partnerships in cybersecurity. Over the past two decades, these collaborations have effectively responded to significant cyber threats, including Distributed Denial of Service (DDoS) attacks and campaigns targeting financial transaction networks. These partnerships have improved the resilience of the financial sector by fostering information sharing, joint response strategies, and regulatory alignment [9]. |
| Global Initiatives | International cooperation is vital for strengthening cybersecurity in the financial sector, especially as cyber threats become more complex and global. Multinational initiatives are instrumental in closing resource gaps and enhancing the resilience of financial institutions, particularly in developing nations. The Carnegie Endowment for International Peace emphasizes that global strategies are essential to address both enduring threats and emerging risks in the cybersecurity landscape. Efforts include cross-border data sharing, capacity building, and joint defense mechanisms [10]. |
| Industry-Led Efforts | Financial institutions, including fintech companies, are central to advancing cybersecurity measures. These organizations are proactive in implementing strategies to counter cyber risks and protect financial systems. Successful industry-led initiatives depend on strong communication and collaboration between all stakeholders, such as regulators, financial institutions, and third-party vendors to minimize risks, safeguard customer data, and maintain the trust of the public. Effective collaboration enhances overall cyber resilience [11]. |
| Case Studies and Surveys | Surveys conducted by organizations like McKinsey and the Institute of International Finance (IIF) offer key insights into global financial institutions' cybersecurity practices. These studies reveal a critical need to enhance cybersecurity in areas such as third-party risk management, privileged access management, and incident response capabilities. Financial |

| Category | Details |
|---|---|
| | institutions continue to face challenges in ensuring comprehensive cybersecurity coverage, especially when dealing with external vendors and emerging threats [12]. |

a. Table 1. An overview of the key areas of collaboration in the financial sector's cybersecurity efforts

*A. Institutional Theory*

Institutional theory looks at how an organization's surroundings affect how it works [7] [13] [14]. Institutions, which are broad frameworks, establish the criteria and conduct considered appropriate and essential for the organization. Over time, these "rules" become deeply integrated as what's expected from businesses [7]. Institutional theory does not directly make a company perform better, but they do make its actions more acceptable in different social settings where businesses operate [15]. Institutional theory looks at how social structures and history shape how organizations, professions, and families form and grow. It explores how the environment affects institutions and the behavior of people within them. Drawing on ideas from social sciences, institutional theory explains how social and cultural surroundings influence both individual and institutional behavior.

Institutional theory has two main approaches. First, it looks at how institutions follow outside influences like government laws or authority rules. Second, it explores the internal processes and routines that institutions develop. Over time, organizations become more formal and rely on impersonal rules and structures. Sometimes, this means they prioritize meeting outside requirements over their own goals. Institutional theory also explains how groups and companies create and stick to norms and standards. They usually follow these rules, making institutions more alike over time, a process called isomorphism. Conforming to industry norms can help push for change within an organization, as initiatives following these norms are more likely to get support. Institutions often adopt ideas that seem "rational" based on their culture and history. These "rationalized myths" can be influenced by outside factors like the government and might not always be truly rational. For example, some schools use teaching methods based on research that might not work well in real classrooms [16].

The study "Collaboration in Financial Sector Cybersecurity: Exploring the Impact of Resource Sharing on IT Security" also uses Institutional Theory to investigate how external factors such as social norms and regulations influence the collaboration among financial institutions regarding cybersecurity. Financial institutions operate within a framework that encompasses regulations from governing bodies, industry norms, and expectations from peers. These external pressures significantly shape how financial institutions cooperate to address cybersecurity challenges. Regulatory requirements, such as compliance standards and data protection laws, often drive financial institutions to collaborate on cybersecurity initiatives. Institutional Theory offers insights into how these external factors influence the dynamics of collaboration in cybersecurity within the financial sector. It shows us how these forces affect whether financial institutions share resources and how it impacts their IT security [17].

*B. Protection Motivation Theory*

Protection motivation theory is a psychological and sociological concept that explains why and how people are motivated to change their behavior. It focuses on the risks or perceived risks that can lead to these changes and the factors required for individuals to follow through with them. Since its inception by Rogers, the theory's application has broadened. Fear appeals, which aim to motivate people through the fear of potential consequences, are utilized in various situations. While some applications aim to benefit individuals or groups, others may be more manipulative in nature [18].

The essence of PMT is centered around how individuals evaluate both the nature of the threat and their capacity to deal with it. PMT has been applied to organizational contexts, particularly in understanding employee behavior related to security compliance. Research frameworks based on PMT have explored variables like cybersecurity leadership, training frequency, and government alerts [19]. The significance of the two-part evaluation procedure cannot be overstated. Firstly, the threat appraisal entails assessing both the gravity and vulnerability of the possible hazard. Severity pertains to the gravity of the outcomes, comprising the physical, emotional, social, and economic ramifications. When considering the risks of smoking, one should reflect on the gravity of lung cancer, considering the potentially catastrophic impact it could have on one's well-being and financial situation. Susceptibility, however, pertains to the probability of someone encountering the threat. Variables such as individual genetic background and levels of exposure to environmental factors are influential. An individual with a familial predisposition to heart disease may experience heightened vulnerability to cardiovascular issues, leading them to contemplate strategies for reducing that risk. After assessing the danger, the person proceeds to a coping assessment, where they evaluate their capacity to handle the circumstance. Two crucial factors that are considered are response efficacy and self-efficacy. Response efficacy refers to the extent to which the prescribed activities are effective in minimizing the threat. An individual contemplating stopping smoking may conduct research on the efficacy of smoking cessation programs to assess their success rates. Self-efficacy pertains to an individual's confidence and capability to effectively carry out specific behaviors. The smoker should contemplate previous endeavors to quit and evaluate their level of assurance in achieving success on this occasion.

Higher perceptions of severity and vulnerability increase motivation for protective actions. Another one is coping appraisal, which evaluates an individual's perceived ability to cope with the threat. It considers their beliefs about the effectiveness of recommended protective behaviors (response efficacy) and their confidence in performing those behaviors (self-efficacy). Higher perceptions of response efficacy and self-efficacy also enhance motivation for protective behaviors. The theory says we're more likely to protect ourselves when we see the danger as serious and likely and when we believe the plan will work and we can do it. But if we don't think the danger is that serious or likely, or we doubt the plan will work or we can do it, we might not act or might even ignore the danger. In real life, this theory helps us understand why people do things to stay healthy, like quitting smoking or getting vaccinated. It also helps make health messages more effective by focusing on making people understand how serious the danger is and how likely it is and making them feel confident they can do what's needed to stay safe [20].

In the financial sector, understanding the severity of cybersecurity threats (such as data breaches, ransomware attacks, or insider threats) is crucial. By applying PMT, researchers can analyze how financial institutions perceive these threats and evaluate their vulnerability. In the context of resource sharing, financial institutions can assess their ability to cope with cybersecurity risks more effectively. Employing actionable risk assessment methods offers a straightforward approach to identifying, evaluating, and mitigating cybersecurity risks, enabling institutions to adapt effectively to emerging threats. These straightforward methods reduce too much reliance on external vendors by simplifying the process and making it more accessible [21]. Researchers can investigate how collaborative efforts (such as sharing threat intelligence, best practices, and resources) enhance response efficacy and self-efficacy. Integrating PMT into the research on resource sharing in financial sector cybersecurity provides insights into threat perception, coping strategies, and risk management. By understanding how institutions collaborate and protect themselves, we can enhance the effectiveness of cybersecurity practices.

*C. Symbolic and Substantive Adoption*

Institutional theory says that institutions look more legitimate when they visibly follow society's rules [22]. According to this idea, symbolic actions are used to show that they're conforming to outside pressures without making big changes inside. These actions also help them stay flexible internally [14]. So, companies and their leaders might focus more on symbolic actions than on making real changes [23]. Studies support the idea that symbolic actions are common. For example, managers might agree to new rules to please shareholders but not fully put them into practice [24] [25]. Symbolic adoption is all about using visible actions to shape how people see an organization, aiming to gain benefits like looking more legitimate. This idea matches with impression management, which was first about how individuals present themselves but is also used to understand how organizations respond when their legitimacy is questioned [26] [27] [28]. Substantive adoption means companies put a lot of time and money into making big changes to improve how they work, even if it means giving up some flexibility inside. It's about really committing to and doing a new idea or way of doing things [23]. For instance, it could involve spending a considerable amount on research and development for a long time to create products that are better for the environment.

In the context of financial sector cybersecurity, symbolic adoption could manifest as policy endorsement, like financial institutions publicly expressing support for collaborative efforts and resource sharing to enhance cybersecurity. Substantive adoption involves concrete actions and practical implementation, like financial organizations actively participating in information-sharing platforms, threat intelligence exchanges, and industry forums. Both symbolic and substantive adoption contribute to the effectiveness of collaborative cybersecurity efforts.

III. RESEARCH MODEL AND HYPOTHESIS DEVELOPMENT

Resource sharing has the potential to function as an independent variable. Resource sharing in the cybersecurity realm among financial institutions applies to the extent to which these organizations cooperate and trade different assets, such as threat intelligence, expertise, technology, and best practices, to enhance their overall cybersecurity position. In a world where financial systems are becoming more interconnected and reliant on digital technology, cyber threats are becoming more advanced and frequent. As a result, pooling resources has become a crucial method for reducing cyber risks and improving overall resilience. Resource sharing is a combination of financial organizations' resources and expertise to efficiently tackle cybersecurity challenges. These resources could involve knowledge about emerging risks, analysis of cybersecurity trends, availability of cutting-edge technological tools, and proficiency in cybersecurity protocols and techniques. Financial institutions participate in resource sharing to harness pooled intelligence and skills in order to enhance their protection mechanisms against cyber threats. For example, the Federal Financial Institutions Examination Council (FFIEC) provides Cybersecurity Assessment Tool (CAT), a comprehensive assessment tool for financial institutions to evaluate their cybersecurity preparedness [29].

Collaboration, as a mediating factor, refers to how effectively and efficiently financial institutions work together in the field of cybersecurity. It involves efficient collaboration to address evolving threats and synchronize actions in response to attacks. This cooperative approach includes not only sharing information and resources but also fostering trust, transparent communication, and joint problem-solving. Effective collaboration allows institutions to leverage each other's expertise and resources for more robust cybersecurity. Surveys or interviews can assess participants' attitudes toward collaboration and its impact on their cooperative efforts. Some examples of collaboration in financial cybersecurity are Joint Financial Associations Cybersecurity Summits, Financial Services Information Sharing and Analysis Center (FS-ISAC), Cross-Functional Collaboration within Organizations etc.

Regulatory authority is a way of making decisions in which an individual or group has much influence and authority over important choices in a system or organization. A regulatory authority, often referred to as a regulatory body or agency, is an independent entity established by the government to enforce regulations related to occupational health and safety. Its primary responsibility is to set and enhance standards and ensure their continuous adherence by individuals and organizations in the regulated sector [30]. There are various regulatory agencies in the United States tasked with supervising and managing financial institutions and markets. These include the Federal Reserve Board (FRB), the Federal Deposit Insurance Corp. (FDIC), and the Securities and Exchange Commission (SEC). The regulation of banks in the United States varies, with some being regulated at the federal level, some at the state level, and some by both federal and state authorities, depending on their charter and structure. Regulatory pressure in banking refers to the influence exerted by government bodies, regulators, or authorities on financial institutions to comply with specific rules, standards, and guidelines. These regulatory pressures can significantly impact how banks operate, manage risk, and serve their customers.

According to NIST, a cybersecurity incident is "A cybersecurity event that has been determined to have an impact on the organization, prompting the need for response and recovery. An occurrence that actually or imminently jeopardizes, without lawful authority, the integrity, confidentiality, or availability of information or an information system or constitutes a violation or imminent threat of violation of law, security policies, security procedures, or acceptable use policies" [31]. In financial organizations, several types of cybersecurity incidents occur, such as data breaches, ransomware attacks, phishing, insider threats, Denial-of-Service (DoS) attacks, malware infections, and fraudulent transactions. These incidents can have significant consequences, including financial, reputational, legal, regulatory, and operational disruption.

Trust is an essential element of teamwork. We might investigate the levels of trust among institutions involved in collaborative cybersecurity initiatives, including faith in the integrity and dependability of shared information. Efficient collaboration depends on consistent and transparent communication channels between the institutions involved. Researchers can evaluate the frequency and methods of communication employed by organizations, such as meetings, email exchanges, or encrypted messaging platforms. Collaboration entails the shared process of identifying and resolving cybersecurity challenges. Researchers might investigate the degree to which institutions participate in collaborative problem-solving activities, such as exchanging knowledge about new threats, conducting joint evaluations of vulnerability, or working together on strategies for responding to incidents.

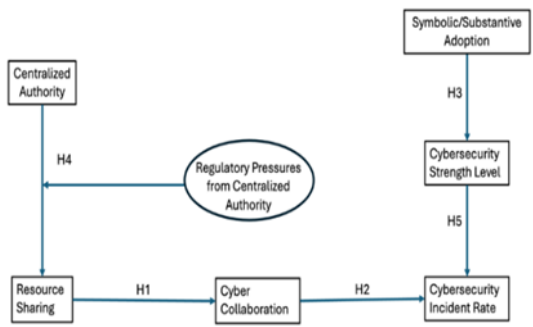

Fig. 1. Conceptual Model of FSCsec

Collaboration and resource sharing are essential for protecting critical infrastructure, including financial institutions. Financial institutions are frequently targeted by hackers because of the vast quantity of sensitive financial data they possess. Attempting to handle cybersecurity on one's own is an unwinnable struggle. Financial institutions can enhance their overall security posture by engaging in resource-sharing and collaboration. Sharing information about recognized risks and vulnerabilities enables institutions to take proactive measures to resolve them before they are maliciously exploited. One way to accomplish this is by utilizing Information Sharing and Analysis Centers (ISACs) that are specifically tailored to the financial industry [32]. Information sharing enables organizations to defend themselves, enhance resilience, and conduct collaborative investigations. Collaboration enables organizations to gain insights and adopt best practices in cybersecurity defense and incident response by leveraging each other's experiences [33]. Frequency of information sharing, participation in collaborative forums, and joint cybersecurity exercises (e.g., red team-blue team exercises) could be the quantitative metrics.

*1) H1: There is a positive relationship between resource sharing and collaboration in financial sector cybersecurity:* Collaborative cybersecurity in financial institutions is the sharing of threat knowledge, optimal methods, and resources to tackle cyber threats proactively and effectively. The use of this collaborative strategy is anticipated to decrease the probability of major security breaches. According to SC Media (2024), there has been an increase in cyberattacks targeting financial institutions. The article highlights the need to foster a security culture that promotes collaboration and resource sharing to enhance resilience against advanced threats [34]. Financial institutions that actively participate in cooperative cybersecurity initiatives are more prepared to protect against cyber-attacks and, as a result, have a lower probability of experiencing significant security breaches. Collaboration enables the consolidation of knowledge, assets, and insight, resulting in a more robust defense against ever-intricate cyber assaults. The hypothesis suggests that when financial institutions actively share resources related to cybersecurity, they are more likely to collaborate effectively. Resource Sharing involves the exchange of information, tools, expertise, and best practices among organizations. It encompasses threat intelligence, incident response strategies, vulnerability assessments, and more. In the context of cybersecurity, collaboration refers to joint efforts, cooperation, and alignment among stakeholders (banks, fintech companies, regulatory bodies, etc.) to enhance security measures. The US Financial Services Sector (FSS) exemplifies effective collaboration through public-private partnerships [9]. Financial institutions work closely with government agencies, sharing threat intelligence, incident data, and best practices.

*2) H2: Financial institutions that actively participate in collaborative cybersecurity are less likely to experience significant cyber breaches than those that do not:* Financial institutions that really implement cybersecurity measures (substantive adoption) and adopt them for regulatory compliance or image (symbolic adoption) would have a more robust defense against cyber threats than those that solely take measures for superficial reasons. They can assess the extent of compliance-driven actions (e.g., policy adherence, regulatory compliance) and also evaluate the depth of genuine cybersecurity investment (e.g., dedicated security teams, advanced tools). Banks that prioritize cybersecurity and fulfill external expectations will have stronger protection compared to banks that merely achieve the minimum requirements to seem compliant. Angst et al. (2017) examine the impact of institutional determinants on the tendency of firms to either symbolically or substantively embrace IT security measures. They find that substantive adoption, influenced by deeper integration of security into IT processes, leads to fewer breaches [2]. According to Frank Cremer, open databases are scarce, impeding collective efforts to manage cyber risks effectively [35]. His findings denote that the lack of accessible data on cyber risk hinders stakeholders' ability to address this multifaceted challenge. Dominik Molitor mentions that with the help of machine learning and text analytics, large textual datasets can be analyzed to shed light on the major scams, fraud, and data breaches that most of the financial institutes have been facing [36]. Another framework is proposed by Fakhrul Safitra to counterattack cyber

disasters by owning cyber resilience and organizations can become more adept in anticipating, averting, responding to, and recovering from cyberattacks. To address the complex dynamics between digitization and cyber security and create a more secure and reliable digital environment going forward, this framework provides strategic direction [37].

*3) H3: Financial institutions that adopt both symbolic and substantive cybersecurity measures have a better overall cybersecurity strength level than those with no involvement:* Financial institutions function within an extensive regulatory framework that includes a range of rules, norms, and standards designed to guarantee the reliability and solidity of the financial system. Institutions are frequently mandated to implement precise cybersecurity procedures to safeguard against cyber threats. Cybersecurity strength level refers to the overall robustness and effectiveness of an organization's cybersecurity measures, including policies, technologies, and practices. Financial organizations are compelled to invest additional money in their cybersecurity operations due to regulatory restrictions. Investments in cybersecurity by private companies, including banks, can be boosted by regulatory mandates, according to an analytical methodology created by [38]. The frequency and quality of resource sharing among financial institutions can be greatly enhanced by establishing a regulatory authority or governing body committed to cybersecurity collaboration. Having a governing body can greatly improve the way financial institutions work together and share resources. Improving the sector's overall cybersecurity posture relies heavily on the central authority's role in setting norms, encouraging communication, and providing a platform for information exchange. Aritran Piplai proposed in one of his papers that Neuro-Symbolic Artificial Intelligence (AI) enhances accessibility and safety in AI systems by combining neural networks' data processing capabilities with symbolic reasoning mechanisms. It identifies the challenges of generating human-understandable explanations for outcomes, especially in domains like cybersecurity and privacy [39]. Anna Himmelhuber suggested that to enhance the usefulness of alerts for analysts, symbolic and sub-symbolic methods such as knowledge graphs can be used to improve the explainability and quality in modern industrial systems. By integrating symbolic reasoning with sub-symbolic data-driven approaches, the research aims to provide actionable insights and facilitate effective responses to cyber threats [40]. While not directly focused on the combination of symbolic and substantive methods, the framework provided by Masike Malatji dedicates valuable insights into the broader AI-driven cybersecurity landscape. Collaborative efforts that integrate symbolic gestures with substantive security practices such as network segmentation and incident response planning contribute to a holistic defense against cyber risks [41].

*4) H4: Regulatory pressures from a centralized authority enhance the frequency of resource sharing among financial institutions:* Organizations with strong cybersecurity measures typically have advanced security infrastructure and well-trained staff. These resources allow them to proactively detect and counter threats, significantly reducing the chances of successful attacks. Additionally, comprehensive cybersecurity programs include employee training on security best practices and proper handling of sensitive data, which is essential for minimizing human error—one of the common causes of cybersecurity breaches. A strong cybersecurity posture is expected to reduce vulnerabilities, enhance threat detection, and improve incident response capabilities. Organizations with better security practices are likely to experience fewer successful cyber breaches [35]. By assessing the incident rate and severity (measuring the impact of each incident as minor, moderate, or severe), we can determine the cybersecurity strength index. David McNulty investigated the possibilities for integrating regulatory goals into financial organizations' management and decision-making processes by utilizing contemporary technologies for data access and sharing. The findings suggest that This can significantly reduce compliance costs and improve oversight of prudential and behavior concerns, and one could see this strategy as an even more extreme advancement of the well-established "new governance" methods of financial regulation [42].

*5) H5: Financial institutions with a higher cybersecurity strength level will experience a lower cybersecurity incident rate compared to institutions with a lower cybersecurity strength level:* Joachim Ulven reviewed studies of cybersecurity's critical risk areas and found that there is still a gap that occurs everyday threat events and hence proposes a generic threat model. To support this, Alimul Haque also worked on a similar institution category. Their findings assess strength levels, emphasizing factors like risk management and technical defenses [43]. According to Frank Cremer and his team contend that one major obstacle facing parties trying to address this issue is the paucity of data on cyber risk that is currently available. Specifically, their work pinpoints a weakness in public databases that jeopardizes group efforts to more effectively mitigate this range of hazards. To better understand, quantify, and manage cyber risks, cybersecurity experts and the insurance sector will benefit from the evaluation and classification of the resultant data [35].

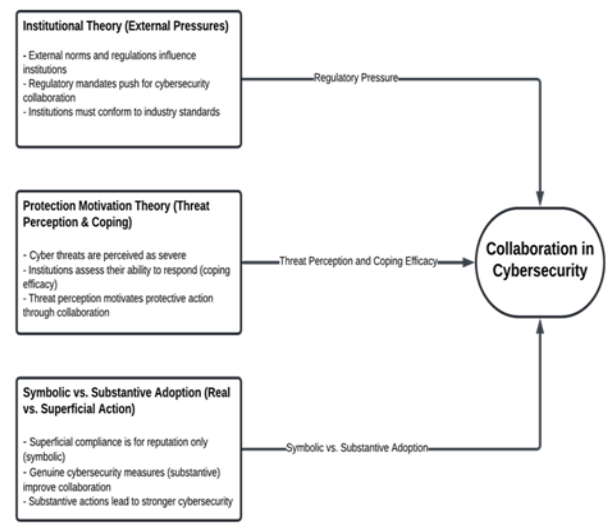

Fig. 2. Collaboration in Financial Sector Cybersecurity

## IV. Discussion

These hypotheses anticipate that banks and financial institutions can improve their cybersecurity by sharing resources and working together. This study thinks that when these institutions collaborate, they can significantly enhance their defenses against cyber-attacks. Financial institutions that share information and resources with each other are better at preventing and managing cyber threats. This is because they can combine their knowledge and technology to build stronger defenses. It's not enough for financial institutions to just agree to share resources; they need to actively participate and implement these strategies to see real benefits. Drawing on various theories, including Institutional Theory, Protection Motivation Theory, and Symbolic and Substantive Adoption, this research shaped hypotheses. These theories provided valuable insights into how external pressures, threat perceptions, and organizational behaviors influence cybersecurity practices. It found that collaboration and resource sharing among financial institutions significantly enhance their cybersecurity posture, aligning with the predictions of Institutional Theory. This theory suggests that institutions conform to external pressures, such as regulatory mandates, by adopting collaborative cybersecurity measures. Additionally, the Protection Motivation Theory helped to understand how threat perceptions influence cybersecurity behaviors. Financial institutions are more motivated to collaborate when they perceive cyber threats as severe and themselves as vulnerable, aligning with the principles of this theory. Moreover, Symbolic and Substantive Adoption theories guided the hypothesis development regarding the adoption of cybersecurity measures. This study posited that institutions engaging in both symbolic and substantive adoption would have a higher cybersecurity strength level than those only adopting symbolic measures. This hypothesis reflects the idea that genuine commitment to cybersecurity practices (substantive adoption) coupled with compliance-driven actions (symbolic adoption) leads to better cybersecurity outcomes.

## V. Conclusions And Future Directions

This study acknowledges the rising and evolving nature of cybersecurity risks to the financial sector. It highlights the importance of active collaboration and resource sharing among financial institutions to enhance cybersecurity. The financial industry requires secure, robust, and reliable systems to ensure uninterrupted operations and maintain public trust. This research highlights the importance of collaboration and resource sharing in improving cybersecurity for financial institutions. By working together, banks can strengthen their defenses against cyber threats and reduce the risk of data breaches. Overall, this research suggests that collaboration and resource sharing are essential strategies for enhancing cybersecurity in the financial sector, ultimately protecting both banks and their customers from cyber-attacks.

In the future scope of this research, several promising areas for further investigation will emerge. Conducting empirical analysis of the hypotheses represents the most crucial future direction for this research. One area is the expansion of collaboration models, examining how partnerships between financial institutions, fintech firms, and regulatory bodies could enhance cybersecurity efforts. Additionally, longitudinal studies could assess the long-term effects of resource sharing on reducing breaches and strengthening institutional resilience. As emerging cyber threats like AI-driven attacks and quantum computing become more prevalent, future research could explore how financial institutions can mitigate these novel risks. A global comparative analysis of regulatory frameworks could provide insights into how varying international regulations impact cybersecurity collaboration. Finally, investigating the potential of advanced technologies such as blockchain, AI, and quantum encryption to bolster resource-sharing strategies and safeguard against sophisticated threats would be a valuable contribution to the field. These directions would build on the current research, broadening the understanding of collaborative cybersecurity strategies in the financial sector.